\documentclass[prb,preprint,showpacs]{revtex4}
\usepackage{amsmath}
\usepackage{amsfonts}
\usepackage{graphicx}

\begin{document}

\title{Optical properties of metal nanoparticles with no center of inversion
symmetry: observation of volume plasmons}
\author{Maxim Sukharev}
\affiliation{Department of Chemistry, Northwestern University, 2145 Sheridan Road,
Evanston, IL. 60208-3113 USA}
\author{Jiha Sung}
\affiliation{Department of Chemistry, Northwestern University, 2145 Sheridan Road,
Evanston, IL. 60208-3113 USA}
\author{Kenneth G. Spears}
\affiliation{Department of Chemistry, Northwestern University, 2145 Sheridan Road,
Evanston, IL. 60208-3113 USA}
\author{Tamar Seideman}
\affiliation{Department of Chemistry, Northwestern University, 2145 Sheridan Road,
Evanston, IL. 60208-3113 USA}
\email{t-seideman@northwestern.edu}

\begin{abstract}
We present theoretical and experimental studies of the optical response of
L-shaped silver nanoparticles. The scattering spectrum exhibits several
plasmon resonances that depend sensitively on the polarization of the
incident electromagnetic field. The physical origin of the resonances is
traced to different plasmon phenomena. In particular, a high energy band
with unusual properties is interpreted in terms of volume plasmon
oscillations arising from the asymmetry of a nanoparticle.
\end{abstract}

\maketitle

\section{Introduction}

The fast progress in fabrication methods \cite{Hutter2004,Slusher1999} of
metallic nanoparticles (MNP) and their arrays, along with advances in laser
technology \cite{Slusher1999} has enabled a wide variety of applications of
nanostructured materials in medicine \cite{VanDuyne2004}, biology \cite%
{Alivisatos2004}, chemistry \cite{VanDuyne2005}, and applied physics \cite%
{Atwater2005}. The recent literature suggests further potential applications
of MNP arrays to guide light in the nanoscale and to study fundamental
questions in wave propagation in confined space \cite{SukharevJPB2007}. The
optical properties of MNPs, and their sensitivity to the particle's shape,
relative arrangement, and surrounding medium \cite{NovotnyBook2006}, result
from strong enhancement of an incident field at the plasmon resonance
frequency \cite{MaierBook2007}, where the light frequency matches the
frequency of collective oscillations of the conduction electrons in the
particle. MNPs with different shapes exhibit a variety of plasmon resonances
ranging from dipole resonances in the quasistatic (diameter$\ll $wavelength)
single sphere limit \cite{KreibigBook}, to quadrupole and multipole modes in
larger spheres \cite{ChumanovJACS2005}, nanocubes \cite{Fuchs1975} and other
structures \cite{Noguez2007}. Particles of spheroidal shapes have been also
a subject of intensive research \cite{HulstBook}, showing multiple-peak
scattering spectra \cite{Aussenegg1996}. Recently it was suggested that MNPs
without a center of inversion symmetry may be used for efficient second
harmonic generation (SHG) \cite{Kauranen2002}.

The tremendous progress in the fabrication of MNPs and analysis of
their properties notwithstanding, several fundamental aspects of
plasmonics remain unclear. In particular, recent experiments on
silver L-shaped nanoparticles and their arrays
\cite{KenJihaExperiment} raise questions on the physical origin of
the multiband scattering spectrum and its dependence on the external
field polarization. In this contribution we present a systematic
study of the optical response of a single L-shaped MNP and
scrutinize the behavior of electromagnetic near fields at the
resonant frequencies in order to explain the observed plasmon
resonances and their physical nature. First, we discuss the theory
and its numerical implementation. Next, we compare our simulations
with experimental measurements. Finally, we discuss the physical
content of each of the plasmon resonances of the L-particle and
introduce the volume plasmon mode as a possible interpretation of
the high energy band observed in experiments.

\section{Theory}

The interaction of metallic nanoparticles with electromagnetic (EM)
radiation is simulated using a finite-difference time-domain approach (FDTD)
\cite{TafloveBook} in three dimensions. The optical response of the metal is
modeled using the auxiliary differential equation method \cite{Taflove1991}
and the standard Drude model, within which the dielectric constant given as%
\begin{equation}
\varepsilon \left( \omega \right) =\varepsilon _{\infty }-\frac{\omega
_{p}^{2}}{\omega ^{2}+i\Gamma \omega },  \label{Drude model}
\end{equation}%
where $\varepsilon _{\infty }$ is the dimensionless infinite-frequency limit
of the dielectric constant, $\omega _{p}$ is the bulk plasma frequency, and $%
\Gamma $ is a damping constant. In order to simulate an open system, we
implement perfectly matched layers (PML) absorbing boundaries with the
exponentially differenced equations, so as to avoid diffusion instabilities
\cite{SadikuBook}. As an incident light source we use a linearly polarized
plane wave, which is numerically generated at each point of a chosen $xy$%
-plane placed above a metal particle and located near the upper PML
region. In order to produce a pure plane wave and avoid the
accumulation of spurious electric charges at the boundaries of the
excitation plane, we embed its ends in PML regions. The parameters
used in our simulations are provided in endnote \cite{Parameters}.

The optical response is simulated using a short-pulse excitation
procedure described elsewhere \cite{GreyPRB2003,Sukharev2007}.
First, we excite the particle by an ultra-short, linearly polarized
pulse (the excitation pulse need be sufficiently short to span the
range of frequencies of relevance; here we used pulses of $0.36$ fs
duration). Next, we propagate the Maxwell equations in time for ca
$60$ fs, recording the electric field components as functions of
time at a fixed point in space. Finally, we perform a fast Fourier
transformation to generate the spectra. The main advantage of this
procedure is that one is able to recover the optical response of a
given structure in terms of different electric field components. As
shown below, it enables us to analyze quantitatively the behavior of
the electric field induced by the asymmetric nanoparticles. Likewise
informative are the vectorial eigenmodes of the system. The latter
can be visualized using phasor functions \cite{TafloveBook}, which
represent steady-state solutions of Maxwell's equations for CW
excitation at the frequency corresponding to
one of eigenenergies of the particle,%
\begin{equation}
\breve{E}_{n}\left( \vec{r};\omega \right) =\rho _{n}\left( \vec{r};\omega
\right) \exp \left( i\phi _{n}\left( \vec{r};\omega \right) \right) ,
\label{phasor}
\end{equation}%
where $n=x,y,z$, and $\rho _{n}$ and $\phi _{n}$ denote the amplitude and
phase of the phasor, respectively. The phasors in Eq. (\ref%
{phasor}) are computed via a recursive discrete Fourier transform
"on the fly" within a single FDTD run \cite{TafloveBook}. In order
to obtain the correct phasor functions (i.e. their amplitudes and
phases) one has to choose carefully the propagation time of
electromagnetic field . In numerical experiments we found that for a
plane wave incident source with a time dependence of the $\sin
\left( \omega t\right) $ form ($\omega $ corresponding to the one of
eigenfrequencies of the structure), the calculation is numerically
converged with $100$ fs propagation time.

In this work we consider the optical properties of L-shaped silver
nanoparticles, as depicted schematically in Fig. 1. All simulations are
performed on the BlueGene/L supercomputer at Argonne National Laboratories.
In order to partition the FDTD scheme onto a parallel grid, we divide the
simulation cube into $64$ slices along the $z$-axis, see Fig. 1, and
implement point-to-point MPI communication subroutines at the boundaries
between slices. The number of $xy$-planes in each slice varies from $5$ to $%
10$. For an FDTD cube of size $L_{x}\times L_{y}\times L_{z}=264\times
264\times 384$ nm and a step size $\delta =1.2$ nm, a single run that
propagates the EM fields for $60$ fs takes on average ca $80$ minutes on $64$
processors.

\section{Results and discussion}

We begin this section by comparing the FDTD simulations described in
the previous section with measurements of scattering and extinction
spectra of a single silver L-particle for different polarizations of
the incident field. Figure 2 presents both the experimental data and
the numerical simulations. The latter involve each $10$ independent
FDTD runs, in which we calculated the scattering intensity along the
incident field polarization for an L-particle in vacuum, taking
$\alpha _{\text{inc}}=\pi /4$ and $3\pi /4$ (see Fig. 1). The
particle dimensions are $H=30$ nm, $L=145$ nm, $d=60$ nm. In order
to estimate the substrate effects we introduce an effective
refractive index, $n_{\text{eff}}$.  By
comparing experimental and simulated resonant wavelengths for the blue ($%
\alpha _{\text{inc}}=\pi /4$) and red ($\alpha _{\text{inc}}=3\pi
/4$) bands we found that $n_{\text{eff}}=1.3$ best matches the
experimental observations. Random deviations of the particle
dimensions are accounted for by performing simulations for a small
range of particle sizes and averaging the resulting spectra over the
ensemble of different sized particle.

The experimental particles have random deviations in the particle arm
length, $L$, and its thickness, $d$. A previous publication \cite%
{KenJihaExperiment} gives a detailed description of the sample fabrication
and measurement setup. In brief, the two-dimensional square array of
L-shaped silver nanoparticles was fabricated by electron-beam lithography on
an indium-tin oxide (ITO) conducting layer of $40$ nm on $750$ $\mu $m thick
glass substrate. The grid spacing of the nanoparticle array is $5$ $\mu $m.
The dipole interaction between particles is negligible at this grid spacing,
therefore each particle can be considered as an isolated particle. The
particle arm length, $L$, thickness, $d$, and height, $H$, are $145$ $\pm $ $%
8$ nm, $63$ $\pm $ $4$ nm, and $30$ nm, respectively. The tips of
each nanoparticle are rounded, as seen in scanning electron
microscope (SEM) images (not shown). The blue band (C) and the high
energy shoulder of the red band (D, $400-800$ nm range) in Fig. 2
are dark-field scattering spectra of a selected nanoparticle within
the sample with $\alpha _{\text{inc}}=\pi /4$ and $3\pi /4$,
respectively. They are obtained on an inverted microscope equipped
with a dark-field condenser and a spectrometer with a liquid
nitrogen-cooled charge-coupled device detector. The red band (D,
$900-1400$ nm
range) is the extinction spectrum of the sample with $\alpha _{\text{inc}%
}=3\pi /4$, obtained with a laboratory-built microscope that records
polarized extinction spectra over a spot diameter (FWHM) of $20$ $\mu $m.

Clearly, the optical response of the L-particle is very sensitive to the
incident field polarization and exhibits four resonant bands. We note that
if the incident field polarization is not along one of two axes of symmetry (%
$\alpha _{\text{inc}}\neq \pi /4,3\pi /4$, see Fig. 1) the spectrum exhibits
all the resonance features, with the amplitudes depending on $\alpha _{\text{%
inc}}$. The physical origin of the blue (C) and red (D) bands can be
understood through extension of the simple case of the eigenmodes of a
simple metallic wire. Collective oscillations of conductive electrons along
the wire axis leads to a low energy resonance, whereas oscillations
perpendicular to the wire axis give rise to a higher energy band. For the
case of a more complex structure, such as the L-structure considered here,
the concept of a shape functional \cite{Fuchs1975,Karam1997} provides an
analogous picture. The shape functionals determine the polarizability tensor
of the particle and hence the response of the confined electrons to an
external field. The inset of Fig. 3a describes schematically the
construction of a shape functional. Consider an L-shaped two dimensional box
driven into oscillations along the $x$-axis (corresponding to an L-particle
exposed to an $x$-polarized external electric field). The resonant
wavelength of these oscillations is proportional to the average dimension, $%
\left\langle x\right\rangle $,
\begin{equation}
\left\langle x\right\rangle =\frac{1}{h}\int L\left( y\right) dy,
\label{averaged L}
\end{equation}%
and depends on the orientation of the particle with respect to the $x$-axis
(see the inset of Fig. 3a). The dependence of $\left\langle x\right\rangle $
on the orientation, shown in the main panel of Fig. 3a, exhibits two
extrema, at $\alpha _{\text{inc}}=\pi /4,3\pi /4$. The two low energy bands
become degenerate in the limit of a parallelepiped shape, as illustrated in
the main panel of Fig. 3b. Here we show the resonant wavelengths of the C
and D bands as functions of the thickness, $d$, of the L-particle arm
starting from $d=20$ nm and ending with the parallelepiped case, $d=150$ nm.
Whereas the energy of the red band (marked D) monotonically increases, that
of the blue band (C) is nonmonotonic, shifting to the blue at small $d$ and
to the red subsequently. The latter corresponds to the case where the high
energy bands A and B (see Fig. 2) strongly overlap with the blue band,
resulting in broad spikes with a fine multi-peak structure.

The origin of the resonance labeled B in Fig. 2 is the sharp corners of the
L-particle. It is well known that sharp features of metallic structures tend
to accumulate surface charges and hence lead to strong EM field enhancements
\cite{NovotnyBook2006}. Figure 4 illustrates the vector field distribution
of phasor functions (\ref{phasor}), $\breve{E}_{x}\left( \vec{r};\omega
\right) $, $\breve{E}_{y}\left( \vec{r};\omega \right) $, in the $xy$-plane
at a distance of $3$ nm above the particle at the wavelength of the
resonance labeled B. It is seen that the corners enhance the EM field and
that each corner gives rise to an oscillating dipole. In order to illustrate
that the origin of the B resonance is the sharp corners, we perform
simulations for an L-shaped particle with rounded corners. The inset of Fig.
4 compares the scattering intensity due to particles with sharp and with
rounded corners, showing that for the latter the B resonance disappears. We
note that a small blue shift of the blue band, C, for the particle with
rounded corners is due to the fact that its volume in our simulations is
slightly smaller than the volume of the L-particle with sharp corners.

A particularly interesting result of this work is the observation of the
high energy band, labeled A Fig. 2. Its origin differs markedly from that of
the features discussed above. Whereas the EM fields corresponding to
resonances B, C, and D are mostly localized on the surface of the particle
and can be readily explained in terms of surface plasmon modes, the EM field
distributions in the case of the A resonance reveal complex volume
oscillations. Figure 5 shows the $xy$-distribution of EM energy calculated
using the phasor function representation (\ref{phasor}) at the middle of the
particle for all four bands. It is evident that the EM energy is localized
in the interior of the particle at the energy of band A. By contrast,
similar plots at the energies of the remaining bands (not shown) emphasize
the surface character of the response. Figure 5a also shows remarkably large
fields inside the particle, reaching almost $3$ order of magnitude
enhancements throughout the entire volume. Additional simulations, in which
we varied the height of the particle, $H$, illustrate strong sensitivity of
the A resonance to the volume of the particle, whereas the lower energy
bands are essentially invariant to the volume. We also analyzed the density
currents for all resonances and found that surface currents dominate the B,
C, and D bands, whereas volume current in the particle plays a dominant role
in the high energy band A. Although the behavior of the density currents at
the energy of band A is similar to that of multipole oscillations, they
differ qualitatively from the latter case in that (by contrast to the
multipole modes \cite{Fuchs1975}) they are not localized on the surface.

By numerically bending a metallic wire and calculating the scattering
spectra as a function of the curvature of such arcuate particle, we verified
that it is breakdown of the inversion symmetry that gives rise to the volume
plasmon modes, which are dipole forbidden for the symmetric particles.
Another interesting observation of our simulations, which distinguishes the
high energy band from the lower energy ones, is the number of oscillations
per resonance lifetime, estimated as the ratio of the eigenenergy of the
plasmon mode to its full width at half maximum \cite{Scully2005}. We found
that this ratio is about $3$-$4$ for the C and D resonances, depending on
the incident polarization, but noticeably larger (ca 8) for the volume
plasmon band. It is interesting to note that a similar high energy band has
been observed in recent experiments on crescent-shaped nanoparticles \cite%
{Kreiter2007}.

\section{Conclusion}

In the previous sections we presented experimental measurements and
numerical analysis of the optical properties of a single L-shaped
silver nanoparticle. We showed that in addition to several
resonances that are attributed to surface plasmon modes, L-particles
also support long-lived volume plasmon oscillations of conductive
electrons, resulting in strong EM fields confined in the interior of
the particle. Our numerical analysis suggests that volume plasmon
modes can be found in spectra of other metallic nanoparticles
without center of inversion symmetry. Among several extensions and
applications of the present work, we plan to explore the optical
response of arrays of asymmetric particles.

\noindent\textbf{Acknowledgments}

This research was supported in part by the NCLT program of the National
Science Foundation (ESI-0426328) at the Materials Research Institute of
Northwestern University, and in part by the Air Force Office of Scientific
Research (MURI program grant F49620-02-1-0381). The numerical work used
resources of the Argonne Leadership Computing Facility at Argonne National
Laboratory, which is supported by the Office of Science of the U.S.
Department of Energy under contract DE-AC02-06CH11357.

\newpage

\begin{figure}[tbph]
\centering\includegraphics[width=\linewidth]{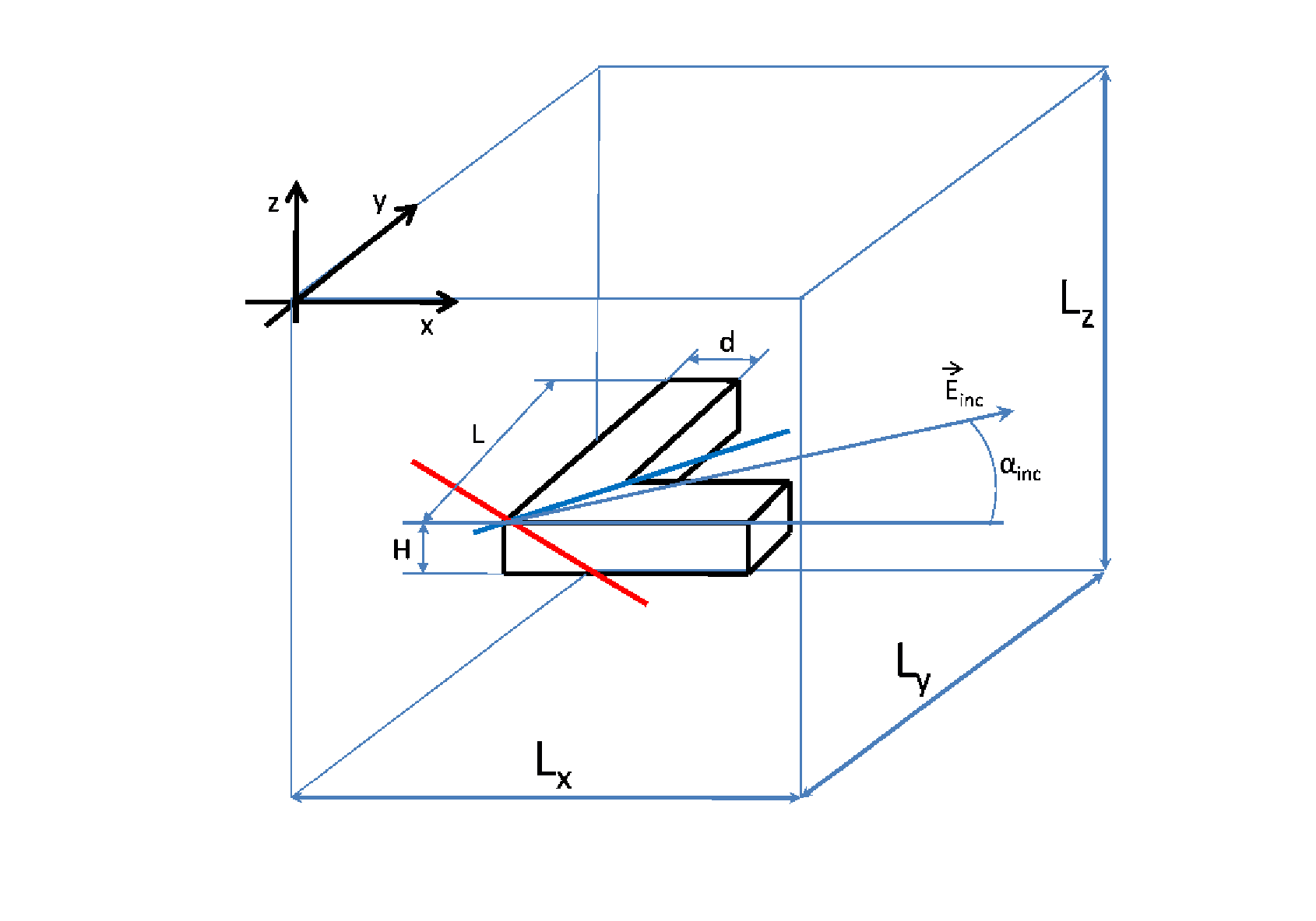}
\caption{(Color online). Schematic setup of the FDTD simulations. L$_{\text{%
x,y,z}}$ denotes the size of the simulation cube. $L$, $d$, and $H$
represent the arm length and thickness, and the height of the L-particle
respectively, here $L=150$ nm, $d=60$ nm, and $H=30$ nm. The particle is
excited by an EM plane wave that is generated $5$ steps below the upper $xy$%
-PML region and is propagated along the $z$-direction. The parameter $%
\protect\alpha _{\text{inc}}$ defines the relative orientation of the
particle with respect to the incident EM field. The red and blue lines
represent the two axes of symmetry of the particle. }
\label{fig1}
\end{figure}

\newpage

\begin{figure}[tbph]
\centering\includegraphics[width=\linewidth]{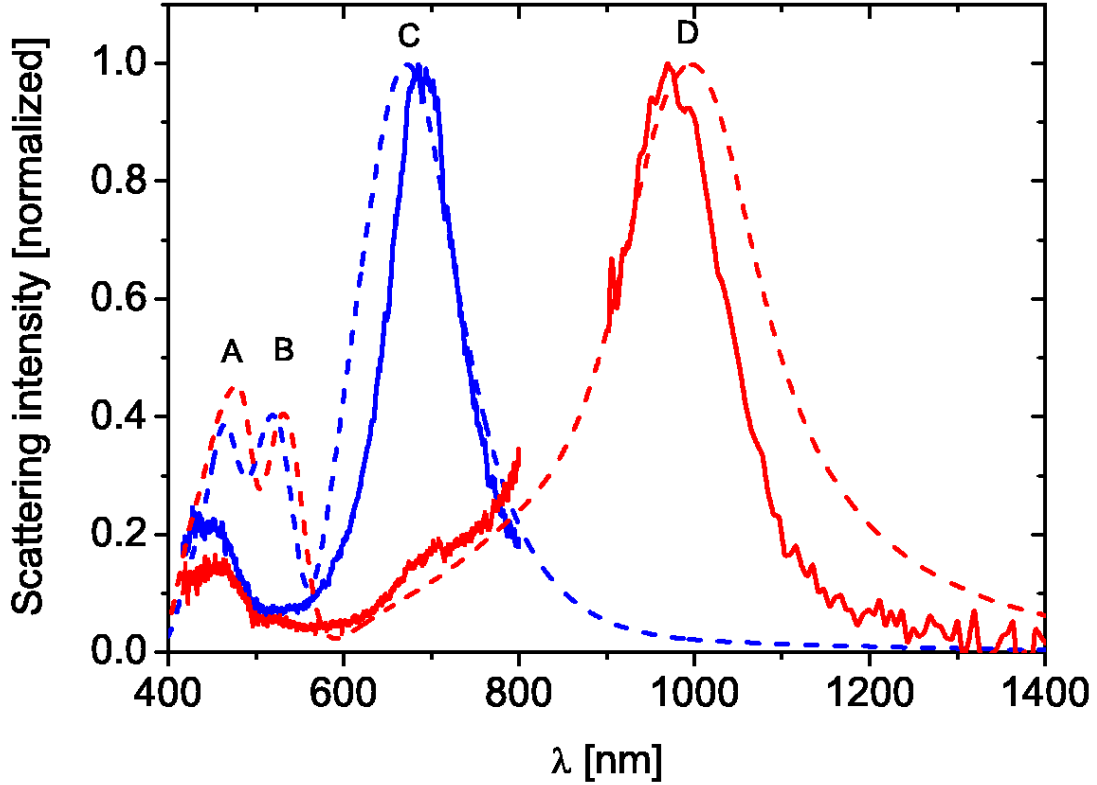}
\caption{(Color online). Normalized experimental (solid curves)
and theoretical (dashed curves) scattering intensity as functions
of the
incident wavelength, $\protect\lambda $. The blue curves correspond to $%
\protect\alpha _{\text{inc}}=\protect\pi /4$ (the incident field
polarization is along the blue axis of symmetry in Fig. 1). The red curves
correspond to $\protect\alpha _{\text{inc}}=3\protect\pi /4$ (the incident
field polarization is along the red axis of symmetry in Fig. 1). A, B, C,
and D indicate the four resonances that are discussed in the text.
Simulations are performed for a particle embedded in nondispersive media
with refractive index $n_{\text{eff}}=1.3$. The small peak near $700$ nm
that can be seen at the shoulder of red band (solid red line) is due to
polarization impurity in the dark field scattering setup. This band was not
observed with pure polarization in two-dimensional array extinction
measurements \protect\cite{KenJihaExperiment}.}
\label{fig2}
\end{figure}

\newpage

\begin{figure}[tbph]
\centering\includegraphics[width=\linewidth]{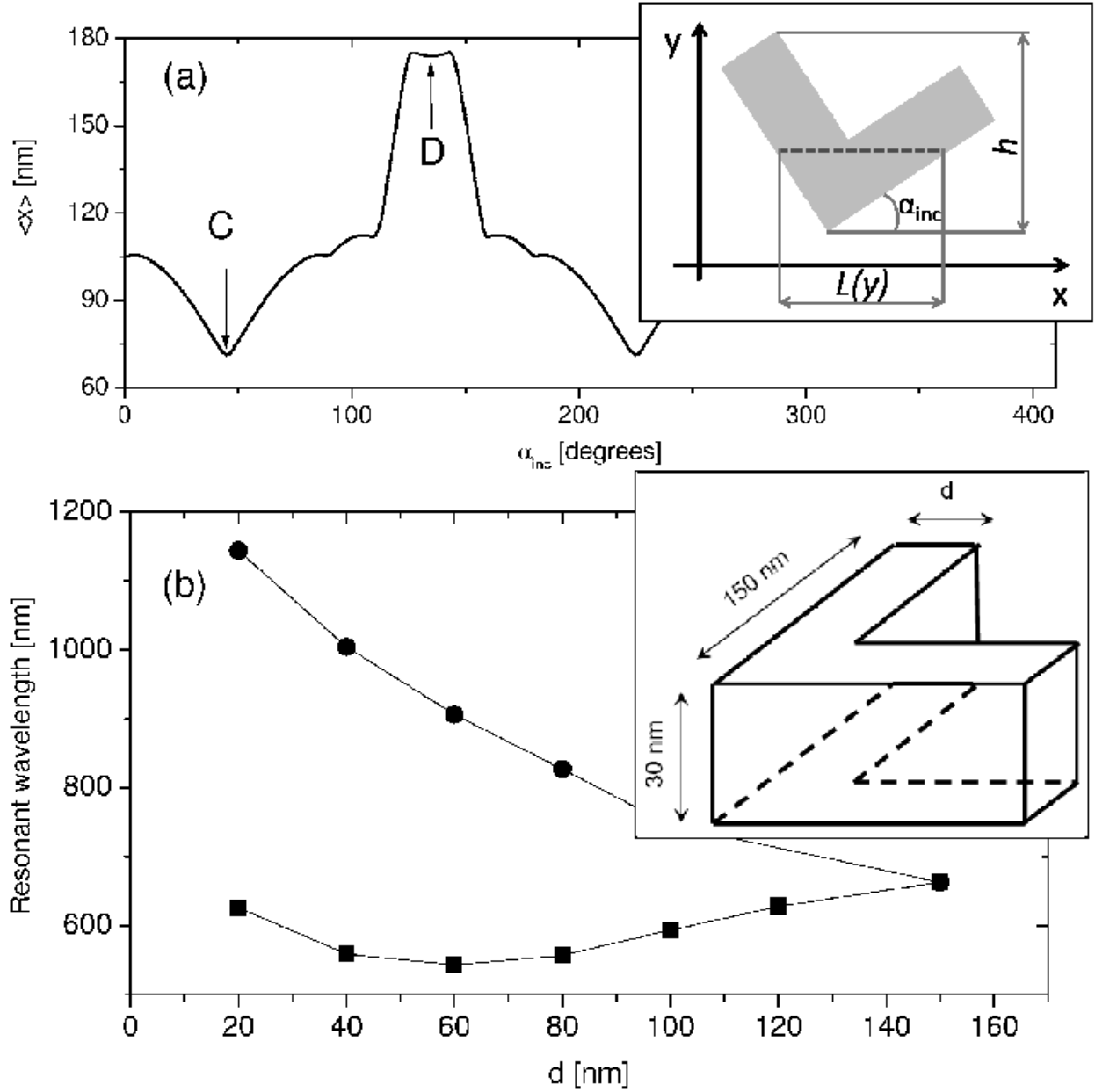}
\caption{(Color online)(a)The concept of a shape functional - the
average length of longitudinal plasmon oscillations, $\left\langle
x\right\rangle $ (see the
upper inset), as a function of the orientation of L-box, $\protect\alpha _{%
\text{inc}}$. (b) The blue and red band positions (i.e., the C and D
resonances in Fig. 2) vs the particle's arm thickness, $d$. Simulations are
performed for a particle embedded in a nondispersive media with refractive
index $n_{\text{eff}}=1.3$.}
\label{fig3}
\end{figure}

\newpage

\begin{figure}[tbph]
\centering\includegraphics[width=\linewidth]{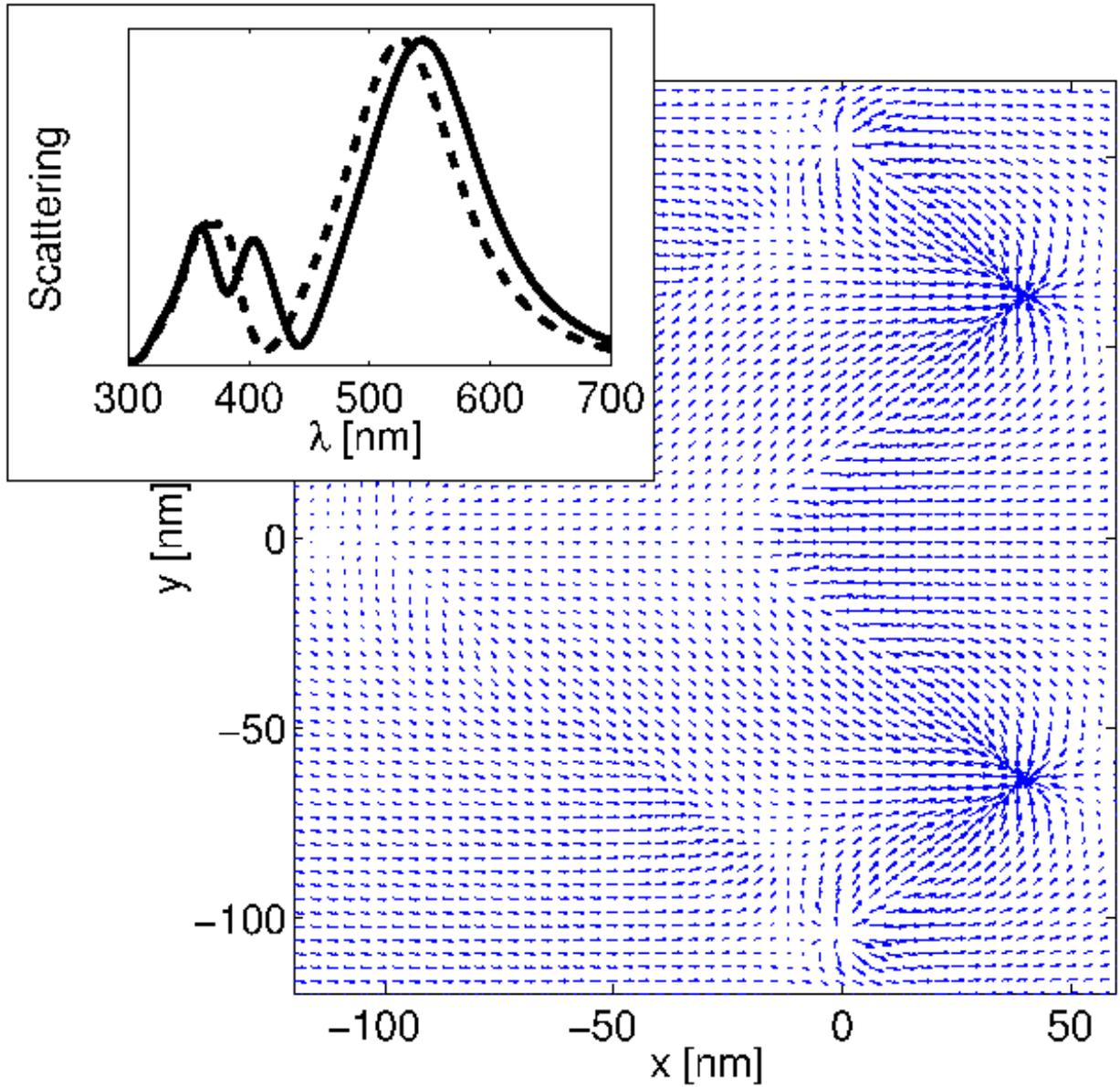}
\caption{Inset: comparison of the scattered intensities as a
function of the incident wavelength, $\protect\lambda $, for
L-particles with sharp (solid curve) and with rounded (dashed
curve) corners. The main panel shows the vector field
$xy$-distribution of electric field inside the particle for the B
resonance (see Fig. 2). Simulations are performed for a particle
in vacuum. } \label{fig4}
\end{figure}

\newpage

\begin{figure}[tbph]
\centering\includegraphics[width=\linewidth]{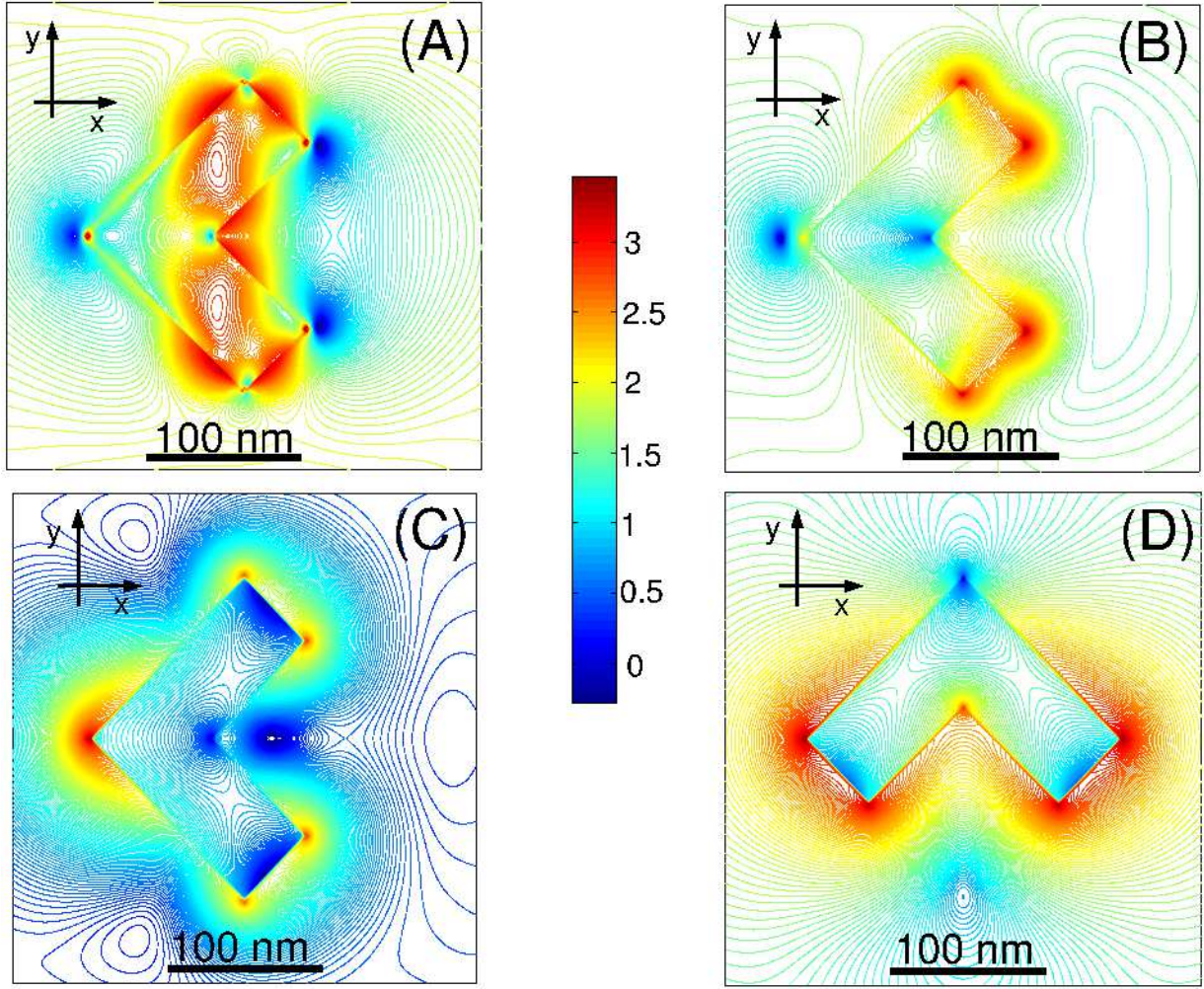}
\caption{(Color online) EM energy enhancement (the ratio of the
total and incident energy) distributions of the plasmon eigenmodes
in the L-particle at $z=0$ (corresponding to the plane bisecting
the particle) on a logarithmic scale. Panel A corresponds to the A
resonance, panel B to the B resonance, panels C to the blue band
and panel D to the red band. The
external EM field is polarized along the $x$-axis with the propagation $k$%
-vector perpendicular to the plane of the figure.}
\label{fig5}
\end{figure}


\begin{thebibliography}{99}
\bibitem{Parameters} Drude parameters in (\ref{Drude model}) for silver used
in our simulations are $\varepsilon _{\infty }=8.926$, $\omega
_{p}=1.7601\times 10^{16}$ rad/sec, and $\Gamma =3.0841\times 10^{14}$
rad/sec.
\end{thebibliography}
\end{document}